\documentclass[12pt]{article}
\begin{document}

\bibliographystyle{unsrt}

\begin{center}
{\bf Generalized qubit portrait of the qutrit state density matrix}\\

 V. N. Chernega$^{(1)}$, O. V. Man'ko$^{(1)}$, V. I. Man'ko$^{((1),(2)}$ \\
(1) - Lebedev Physical Institute, 119991, Leninskii prospect, 57, Moscow, Russia\\
(2) - Moscow Institute of Physics and Technology
\end{center}

 \begin{abstract}
 New inequalities for tomographic probability distributions and density matrices of qutrit states are obtained by means of generalization of qubit portrait method. The approach based on the qudit portrait method to get new entropic inequalities is proposed. It can be applied to the case of arbutrary nonnegative hermitian matrices including the density matrices of multipartite qudit states.
 \end{abstract}

 \section{Introduction}
 The qubit portrait of qudit states \cite{Cher1,Lupo,Fillipov} was used to consider the problem of Bell inequalities \cite{Bell,Horn} and their violation \cite{Cirelson}. In fact, the qubit portrait is the application of particular stohastic matrices with the matrix elements equal either to zero or unity to a given probability distribution vector. The obtained new probability vectors contain information on the initial probability vector. These new probability vectors (called "portrait" of the initial probability vector) can be studied from the point of view of their entropic properties including Renyi entropy \cite{Ren}, Shannon entropy \cite{Shan} and Tsallis entropy \cite{Ts}. The entropic and information properties of the probability vectors which belong to the portrait of the initial probability vector characterize this initial vector.

 The aim of this work is to extend the qubit portrait method to study the information properties of density matrices describing the qudit states. We  suggest the construction of "portrait" of the initial density matrices. The portrait density matrices are particular nonnegative density matrices obtained by means of application of some specific nonnegative maps to the initial density matrices. If the initial density matrix is considered as a complex vector the portrait density matrix is obtained from this vector acting on the vector by a matrix with matrix elements equal either zero or unity. In the spirit of the qubit portrait method where we applied the particular stochastic matrices to probability vectors we are going to apply the analogous matrices (matrices of the particular nonnegative maps) to the hermitian nonnegative matrices. To demonstrate the generic properties of the method we consider in the work a concrete example of 3x3 - density matrix of a qutrit state. In this case we obtain new inequalities for the qutrit state density matrix. The entropic and information inequalities for qudit tomograms were studied recently \cite{RitaVIJRLRN342013} in connection with generic inequalities for sets of nonnegative numbers which yield probability vectors. For the sets of complex numbers which are identified with matrix elements of nonnegative hermitian matrices one can also find the corresponding entropic inequalities.

 The paper is organized as follows. In Sec. 2 we review the qubit portrait method. In Sec. 3 we present the new entropic inequalities for the qutrit state density matrices. In Sec.4 we formulate the generic rules of the qudit portrait approach. Perspective and conclusions are presented in Sec.5.

 \section{Qubit portrait}
 Let us consider a probability vector ${\bf P}=(p_1,p_2,p_3)$. The numbers $p_k$ are nonnegative and satisfy the condition $\sum_{k=1}^3p_k=1.$ The vector corresponds to classical analog of qutrit. The portrait of the probability 3-vector is defined as a set of other probability vectors obtained by means of the stohastic matrices, e.g.
 \begin{equation}\label{eq.1}
 {\bf \Pi_{1,2}}=M_{1,2}{\bf P},\quad M_1=\left(\begin{array}{ccc}
                                                   1 & 1 & 0 \\
                                                   0 & 0 & 1 \\
                                                   0 & 0 & 0 \\
                                                 \end{array}\right),
 \quad M_2=\left(\begin{array}{ccc}
               1 & 0 & 1 \\
               0 & 1 & 0 \\
               0 & 0 & 0 \\
             \end{array}
           \right)
 \end{equation}
 has
 \begin{equation}\label{eq.2}
 {\bf \Pi_{1}}=\left(\begin{array}{c}
                p_1+p_2 \\
                p_3\\
                0
              \end{array}\right),\quad {\bf \Pi_{2}}=\left(\begin{array}{c}
                                                p_1+p_3 \\
                                                p_2 \\
                                                0
                                              \end{array}\right).
 \end{equation}
 The form of the stohastic matrices is special in the sense that the matrix elements are equal to either zeros or unity. It was pointed out that there are inequalities for Shannon entropies related to entropy of the initial probability vector ${\bf P}$ and the portrait vectors.

 It means that one always has the inequality
 \begin{equation}\label{eq.2a}
 -\sum_{k=1}^3p_k\ln p_k\geq-\sum_{k=1}^3({\bf \Pi_{1,2}})_k\ln({\bf \Pi_{1,2}})_k
\end{equation}
The analogous inequalities can be obtained for $N$-vectors also as well the inequalities can be written for Renyi and Tsallis entropies \cite{Ren,Ts}.

\section{Portrait of density matrix}
Our goal is to suggest an extension of the qubit portrait method to study the density matrices. For example, one has instead of the probability 3-vector a 3x3-matrix, e.g. a density matrix of qutrit of the form
\begin{equation}\label{eq.3}
\rho=\left(
       \begin{array}{ccc}
         \rho_{11} & \rho_{12} & \rho_{13} \\
         \rho_{21} & \rho_{22} & \rho_{23} \\
         \rho_{31} & \rho_{32} & \rho_{33} \\
       \end{array}
     \right), \quad \mbox{Tr}\rho=1,\quad \rho^\dagger=\rho.
     \end{equation}
The density matrix has nonnegative eigenvalues, i.e. $\rho\geq0$.

We want to suggest some map which provides 3x3-density matrix obtained as a specific positive map of the initial matrix. For example, there are two matrices of the form
\begin{equation}\label{eq.4}
\rho_1=\left(
         \begin{array}{ccc}
           \rho_{11}+\rho_{22} & \rho_{13} & 0 \\
           \rho_{31} & \rho_{33} & 0 \\
           0 & 0 & 0 \\
         \end{array}
       \right), \quad \rho_2=\left(
         \begin{array}{ccc}
           \rho_{11}+\rho_{33} & \rho_{12} & 0 \\
           \rho_{21} & \rho_{22} & 0 \\
           0 & 0 & 0 \\
         \end{array}
       \right).
\end{equation}
One can check that the maps $\rho\rightarrow\rho_1,$ $\rho\rightarrow\rho_2$ are positive. The maps are done using the procedure to present the matrices in the form of vectors. In our case the 3x3 matrix is presented in the form of the complex 9-vector. The portrait is obtained by acting on the 9-vector by the matrix containing the matrix elements which are equal either zero or unity. In this sense the procedure to get the portrait matrix is analogous to the procedure to get the portrait probability vector. Thus we can write
\begin{equation}\label{eq.5}
\rho_1={\cal M}_1\rho,\quad \rho_2={\cal M}_2\rho,
\end{equation}
where ${\cal M}_{1,2}$ mean the linear maps providing the result (\ref{eq.4}). One can get new entropic inequalities for the density matrices analogous to the inequalities for the probability vectors. Thus for von Neumann entropies one get new inequality
\begin{equation}\label{eq.6}
-\mbox{Tr}\rho\ln\rho\leq-\mbox{Tr}\rho_1\ln\rho_1-\mbox{Tr}\rho_2\ln\rho_2.
\end{equation}
This inequality is an analog of subadditivity condition for density matrix of the bipartite system containing e.g. two qubits.

On the other hand the written inequality corresponds to quantum correlations available for the state of one qutrit which is not bipartite system. We will write the inequality in the explicit form as
\begin{eqnarray}
&&-\mbox{Tr}\{\left(
       \begin{array}{ccc}
         \rho_{11} & \rho_{12} & \rho_{13} \\
         \rho_{21} & \rho_{22} & \rho_{23} \\
         \rho_{31} & \rho_{32} & \rho_{33} \\
       \end{array}
     \right)\ln\left(
       \begin{array}{ccc}
         \rho_{11} & \rho_{12} & \rho_{13} \\
         \rho_{21} & \rho_{22} & \rho_{23} \\
         \rho_{31} & \rho_{32} & \rho_{33} \\
       \end{array}
     \right)\}\nonumber\\
&&     \leq-\mbox{Tr}\{\left(
         \begin{array}{ccc}
           \rho_{11}+\rho_{22} & \rho_{13}  \\
           \rho_{31} & \rho_{33}  \\
                    \end{array}
       \right)\ln\left(
         \begin{array}{ccc}
           \rho_{11}+\rho_{22} & \rho_{13}  \\
           \rho_{31} & \rho_{33}  \\
                    \end{array}
       \right)\}\nonumber\\
      &&-\mbox{Tr}\{\left(
         \begin{array}{ccc}
           \rho_{11}+\rho_{33} & \rho_{12} \\
           \rho_{21} & \rho_{22}  \\
                    \end{array}
       \right)\ln\left(
         \begin{array}{ccc}
           \rho_{11}+\rho_{33} & \rho_{12}  \\
           \rho_{21} & \rho_{22} \\
                    \end{array}
       \right)\}.\label{eq.7}
       \end{eqnarray}
       Thus the Shannon entropy of any qutrit state is less then sum of Shannon entropies of the two portrait qubit states.

       There is a quantum information which can be associated to any qutrit state which is equal to difference of the von Neumann entropies involved into subbaditivity condition, i.e.
       \begin{equation}\label{eq.8}
       I_q=-\mbox{Tr}\rho_1\ln\rho_1-\mbox{Tr}\rho_2\ln\rho_2+\mbox{Tr}\rho\ln\rho.
       \end{equation}
       The quantum information associated to one qutrit state is mandatory nonnegative
       \begin{equation}\label{eq.9}
       I_q\geq0
       \end{equation}
       Another entropic inequality for density matrix of qutrit state can be given on the base of nonnegativity of relative entropy. We write it in explicit form as
       \begin{equation}\label{eq.10}
       0\leq\mbox{Tr}\{\left(
         \begin{array}{ccc}
           \rho_{11}+\rho_{22} & \rho_{13}  \\
           \rho_{31} & \rho_{33}  \\
                    \end{array}
       \right)\ln\left(\left(
         \begin{array}{ccc}
           \rho_{11}+\rho_{22} & \rho_{13}  \\
           \rho_{31} & \rho_{33}  \\
                    \end{array}
       \right)\left(
         \begin{array}{ccc}
           \rho_{11}+\rho_{33} & \rho_{12}  \\
           \rho_{21} & \rho_{22} \\
                    \end{array}
       \right)^{-1}\right)\}.
       \end{equation}

\section{Generic scheme to get new inequalities}
We demonstrated the presence of new entropic inequalities for any qutrit state density matrix. These inequalities correspond to specific quantum correlations coded by matrix elements of hermitian nonnegative density matrices of qutrit. In fact such correlations and analogous entropic and information inequalities exist for arbitrary qudit states including multipartite qudit states. The procedure of visualisation of the entropic inequalities is the following. For a given density matrix (either of one qudit state or multipartite qudit state) the matrix is mapped onto the complex vector. Then by acting by the specific positive maps realized by the matrices containing only zero or unity matrix elements one obtains the vectors which are represented as density matrices of other qudit states. These density matrices are qudit portraits of the initial density matrix. The different rules to make the positive maps of the initial density matrix are appropriate to get analogs either of subadditivety condition or strong subadditivity condition. In fact the inequalities corresponding to these conditions do not depend on the composite structure of the qudit systems. They are valid for any multipartite system density matrix including the case of one qudit. The physical sense and possibility to detect the quantum correlations corresponding to the new inequalities for the density matrices of the qudit states need extra clarification. One can remark that the properties of new introduced quantum information of qutrit state has some analogy with properties of quantum discord.

\section{Conclusions}
In the tomographic picture of quantum mechanics the states are identified with the probability distributions (see, e.g. \cite{IbortPhysScr2009,DodonovPhysLet,OlyaJETP,VovaJETPL}). Any density matrix of systems of qudits $\rho$ provides the probability distribution
\begin{equation}\label{eq.11}
<{\bf m}|u\rho u^\dagger|{\bf m}>=w({\bf m},u),
\end{equation}
where $u$ is unitary matrix and $|{\bf m}>$ is an arbitrary basis in the Hilbert space. For example, the basis for multiqudit state can be  determined in terms of tensor product of states $|m_k>$, where $m_k$ is spin projection on corresponding quantization axes. The probability density depends on unitary matrix $u$. In fact the probability $w({\bf m},u)$ is the conditional probability $w({\bf m}|u)$  which can be connected with a joint probability $P({\bf m},u)$ by means of Bayesian formula
\begin{equation}\label{eq.12}
w({\bf m},u)\Phi(u)=P({\bf m},u),
\end{equation}
where $\Phi(u)$ is an arbitrary probability density on the unitary group.

The conditional probability $w({\bf m}|u)$ can be considered as the probability vector. It has the form
\begin{equation}\label{eq.13}
\vec{w}(u)=|uu_0|^2\vec{\rho}.
\end{equation}
Here the vector $\vec{\rho}$ is the vector with the components equal to eigenvalues of the density matrix $\rho$. The unitary matrix $u_o$ has the columns which correspond to normalized eigenvectors of the matrix $\rho$. The notation for the matrix $|a|^2$ means that it has matrix elements $(|a|^2)_{jk}=|a_{jk}|^2$. The connection of the tomogram (\ref{eq.13}) to the density matrix permits to clarify the method of the qudit portrait of the probability vectors and portrait of the density matrix. The minimum on the unitary group of Shannon entropy of the probability vector $\vec{w}(u)$ is known to determine the von Neumann entropy of the density matrix of the qudit state, i.e.
\begin{equation}\label{eq.14}
min(-\sum_m w({\bf m}|u)\ln w({\bf m}|u))=-\mbox{Tr}\rho\ln\rho.
\end{equation}
One can use this property to consider its analog for the cases of the introduced portrait density matrices. This aspect will be studied in another article.

\section*{Acknowledgements}
This work was partially supported by the Russian Foundation for Basic Research under Project No. 11-02-00456-a.

  \end{document}